\documentstyle[amssymb,preprint,aps,psfig]{revtex}

\begin{document}
\title{Quantum limits in the measurement of very small displacements in optical
images }
\author{{\bf C. Fabre, J.B.\ Fouet, A. Ma\^{\i }tre}}
\address{Laboratoire Kastler Brossel, Universit\'{e} Pierre et Marie Curie, Case 74,\\
75252 Paris cedex 05, France}
\maketitle

\begin{abstract}
We consider the problem of the measurement of very small displacements in
the transverse plane of an optical image using a split-photodetector. We
show that the standard quantum limit for such a measurement, equal to the
diffraction limit divided by the square root of the number of photons used
in the measurement, cannot be overcome by using ''ordinary'' single mode
squeezed light. We give the form of possible multimode nonclassical states
of light enabling us to enhance by orders of magnitude the resolution of
such a measurement beyond the standard quantum limit.

\end{abstract}

\smallskip As CCD cameras and photodetector arrays are now widely used to
record optical images, it is important to know to which extent does the
quantum nature of light impose fundamental limits to the quality and
resolution in this kind of optical measurement. Quantum limits in optics
have been already studied in many configurations, but only in the case of
light monitored by large area photodetectors that give an information
integrated over all the transverse plane. On the other hand, subtle quantum
effects occurring in images are now actively studied\cite{review2}, but not,
to the best of our knowledge, for the determination of quantum limits in
transverse measurements. In order to give precise assessments of such
quantum limits, we will focus our attention to a precise problem, namely the
measurement of very small displacements in an optical image. After recalling
the standard quantum limit for the optical resolution in such a
configuration, we will show that one must use multi-transverse modes
nonclassical states of light to improve the resolution beyond the standard
quantum limit, and give examples of such multimode nonclassical states.

According to the Rayleigh criterion \cite{rayleigh,rushforth}, the
resolution in optical images is limited by diffraction. This classical
criterion, based on the capabilities of the human visual system\cite{survey}%
, can be violated when one uses modern photodetectors which are able to
resolve image details much smaller than the size of a diffraction spot. For
such measurements the resolution is limited by the quantum fluctuations of
the light affecting each pixel \cite{goodman,kolobov}.

Let us take an orthonormal basis $u_{i}\left( x,y\right) $ of transverse
modes (where $x$ and $y$ are the coordinates in the transverse plane), and
call $a_{i}$ the corresponding annihilation operators. The photocurrent $%
N_{S}$ measured on a pixel of area $S$, expressed in units of number of
photons recorded during the measurement time, is equal to the integral over $%
S$ of the quantity $\left\langle (E^{(+)})^{\dagger }E^{(+)}\right\rangle $,
where\ $E^{(+)}(x,y)=\sum_{j}a_{j}u_{j}(x,y)$. Let us first assume that the
input beam is described by a {\it single mode quantum state }in the
transverse mode $u_{1}$. A straightforward calculation based on standard
photodetection theory shows that 
\begin{eqnarray}
C_{N_{A}N_{B}} &=&\left\langle N_{A}N_{B}\right\rangle -\left\langle
N_{A}\right\rangle \left\langle N_{B}\right\rangle  \nonumber \\
&=&N_{S_{A}\cap S_{B}}+\frac{\left\langle N_{A}\right\rangle \left\langle
N_{B}\right\rangle }{\left\langle N_{tot}\right\rangle ^{2}}\left( \Delta
N_{tot}^{2}-\left\langle N_{tot}\right\rangle \right)  \label{eq1}
\end{eqnarray}
where $N_{A}$, $N_{B}$ and $N_{S_{A}\cap S_{B}}$ are the photocurrents
measured by photodetectors of areas $S_{A}$, $S_{B}$ and $S_{A}\cap S_{B}$. $%
\left\langle N_{tot}\right\rangle $ and $\Delta N_{tot}^{2}$ are
respectively the mean and the variance of the total photocurrent, measured
by a detector covering the entire transverse plane. If the quantum state
describing the single mode beam is a coherent state, then $%
C_{N_{A}N_{B}}=N_{S_{A}\cap S_{B}}$ : the photocurrent fluctuations are at
the shot noise level on all the pixels, whatever their size and position,
and the fluctuations at two different pixels are uncorrelated. This result
is consistent with the simple picture that a coherent state is 'composed' of
photons which are randomly distributed not only in time, but also in the
transverse plane inside the beam area.

\smallskip Let us consider as an example a configuration which is very often
used for the measurement of small displacements, for example in atomic force
microscopy\cite{atomforce}, ultra-weak absorption measurement\cite{mirage},
or in single molecule tracking in biology\cite{biology} : the photodetector
array is a two-pixel photodetector (''split detector''), delivering two
photocurrents $N_{A}$ and $N_{B}$ proportional to the light intensities
integrated over two halves of the transverse plane ($A$: $x>0$; $B$: $x<0$%
).\ A light beam of intensity $I\left( x,y\right) $ (expressed in photons
per unit area), assumed symmetrical with respect to the pixel boundary, is
incident on the split photodetector. If the beam is initially centered on
the detector, the mean value of the photocurrent difference $%
N_{-}=N_{A}-N_{B}$ is directly proportional to the relative lateral
displacement $D$ of the whole beam with respect to the initial symmetrical
configuration, at least for displacements small compared to the beam size.
The noise affecting this quantity, sometimes called ''position noise'',
limits the accuracy in the measurement of $D$. It has been studied
experimentally and theoretically for various laser beams \cite
{levenson,grangier2}. When the light beam is in a coherent state, the
displacement $D_{sql}$ providing a value of $\left\langle N_{-}\right\rangle 
$ equal to this noise is the standard quantum limit in the measurement of a
small transverse displacement.\ One finds from Eq(\ref{eq1}) 
\begin{equation}
D_{sql}=\frac{\Delta }{\sqrt{<N_{tot}>}}  \label{eq6}
\end{equation}
where $\Delta $ is the beam effective width, which is defined by 
\begin{equation}
\Delta =\frac{<N_{tot}>}{\left( \frac{\partial <N_{-}>}{\partial D}\right)
_{D=0}}=\frac{<N_{tot}>}{2\int_{-\infty }^{+\infty }dyI\left( 0,y\right) }
\label{eq7}
\end{equation}
and which depends on the exact beam shape.$\;$For example, $\Delta
=0\allowbreak .\,63w_{0}$ for a $TEM_{00}$ Gaussian beam of waist $w_{0}.$ $%
\Delta $ is, within some numerical factor of order one, the Rayleigh, or
diffraction, limit for the optical resolution in this specific measurement.
Expressions analog to (\ref{eq6}) have been already obtained\cite{atomforce}%
. $D_{sql}$ can be much smaller than the optical wavelength even with light
beams of moderate intensities.

Let us now replace the single mode coherent beam by a single mode
nonclassical beam, and more precisely by a sub-Poissonian beam, for which $%
\Delta N_{tot}^{2}<\left\langle N_{tot}\right\rangle $, such as those
produced by some semiconductor lasers\cite{subpoisson}.\ Formula (\ref{eq1})
implies that for any pixel of area $S_{A}$%
\begin{equation}
\frac{\Delta N_{A}^{2}}{\left\langle N_{A}\right\rangle }-1=\frac{%
\left\langle N_{A}\right\rangle }{\left\langle N_{tot}\right\rangle }\left( 
\frac{\Delta N_{tot}^{2}}{\left\langle N_{tot}\right\rangle }-1\right)
\label{eq8}
\end{equation}

As $\left\langle N_{A}\right\rangle \leqslant \left\langle
N_{tot}\right\rangle $, this expression shows that the relative noise
reduction with respect to shot noise is smaller when one measures a part of
a single mode beam than on the total beam.\ The degradation in the measured
intensity squeezing is the proportion $\frac{\left\langle N_{A}\right\rangle 
}{\left\langle N_{tot}\right\rangle }$ of the intensity measured in the
partial detection.\ One gets an expression similar to formula (\ref{eq8})
when one determines the effect on squeezing of a lossy medium of intensity
transmission coefficient $T$ (replacing $\frac{\left\langle
N_{S_{A}}\right\rangle }{\left\langle N_{tot}\right\rangle }$ by $T$).\ For
a single mode beam, a partial detection is thus equivalent to a loss
mechanism. This property can be simply pictured by asserting that a single
mode sub-Poissonian beam is composed of photons which are somehow
antibunched in time, but still completely randomly distributed in the
transverse plane, like in a coherent state : transverse randomness is
therefore associated to the single mode nature of the field, and not to its
coherent or quasi-classical nature.\ From (\ref{eq1}), it is easy to show
that the noise on $N_{-}=N_{A}-N_{B}$ does not depend on the quantum state
of the single mode beam used in the experiment, and therefore that the
minimum displacement measurable with a split detector is still $D_{sql}$: we
are led to the conclusion that a{\it \ single mode nonclassical state cannot
improve the transverse resolution beyond the standard quantum limit}.

One must therefore use a {\it transverse multimode state of light},. Let us
first consider a two-mode nonclassical state, spanned on the first two
transverse modes $u_{1}$ and $u_{2}$. In the small fluctuation limit (i.e.
neglecting terms quadratic in the fluctuations), which is valid when one
uses intense beams, assuming that the modes $u_{1}$ and $u_{2}$ are real,
and using the orthonormality and closure relations for the transverse modes,
the photocurrent fluctuation measured on a pixel of area $S_{A}$ can be
shown to be 
\begin{eqnarray}
\Delta N_{A}^{2} &=&\left\langle N_{A}\right\rangle +\left[
\sum_{i=1,2}(C_{a_{i}^{+}a_{i}}\left| A_{i}\right| ^{2}+C_{a_{i}a_{i}}\left(
A_{i}^{*}\right) ^{2})\right.  \nonumber \\
&&+2C_{a_{1}a_{2}^{+}}A_{2}A_{1}^{*}+2C_{a_{1}a_{2}}A_{1}^{*}A_{2}^{*} 
\nonumber \\
&&\left. +\text{{\it complex conjugate }}\right]  \label{eq9}
\end{eqnarray}
where $C_{a_{i}^{+}a_{i}}$ (and analogous quantities) are the correlation
functions of operators $a_{i}^{+}$ and $a_{i}$, and $A_{i}$ ($i=1,2$) is the
overlap integral on the pixel surface 
\begin{equation}
A_{i}=\int \int_{S_{A}}u_{i}\left( x,y\right) \left\langle E^{(+)}\left(
x,y\right) \right\rangle dxdy  \label{eq10}
\end{equation}
If the system is in a two-mode coherent state, all the quantum mean values
are zero in (\ref{eq9}), except $\left\langle N_{A}\right\rangle $: one
finds again the shot noise, like in the single mode case. If the system is
not in a coherent state, it is no longer possible to write expression (\ref
{eq9}) in a form analogous to (\ref{eq8}), reminiscent of a loss mechanism.
One finds therefore that considering a partial photodetection as equivalent
to a loss is not true in general. One can find in \cite{levenson} and \cite
{grangier2} examples of nontrivial noise variations in partial
photodetection, in the case of multimode laser beams with excess noise,
providing useful information on the laser used in the experiment.

Let us now use a two-mode nonclassical state. It is easy to show that the
noise on the intensity difference $N_{A}-N_{B}$ between the two zones is
given by an expression similar to (\ref{eq9}), where $\left\langle
N_{A}\right\rangle $ is replaced by the total number of photons $%
\left\langle N_{tot}\right\rangle $ in the light beam, and where the $A_{i}$
coefficients are replaced by $A_{i}^{\prime }$ given by 
\begin{equation}
A_{i}^{\prime }=\int \int_{x>0}u_{i}\left\langle E^{(+)}\right\rangle
dxdy-\int \int_{x<0}u_{i}\left\langle E^{(+)}\right\rangle dxdy  \label{eq11}
\end{equation}

Let us give now an example of a two-mode state allowing us to reduce by a
large amount the variance in the measurement of the intensity difference
between the two half planes : we consider an even mode $u_{e}$, and an odd
mode $u_{o}$, with respect to the coordinate $x$, and we assume that the
light state consists of a tensor product of a coherent state having a non
zero mean value in mode $u_{e}$, and of a squeezed vacuum in mode $u_{o}$.\
The mean value of the field $\left\langle E^{(+)}(x,y)\right\rangle $ is
then an even function of $x$, which gives a zero mean value for the measured
signal $N_{A}-N_{B}$ when the displacement is zero.\ Using expressions (\ref
{eq9}) and (\ref{eq11}), and assuming that $\left\langle
E^{(+)}(x,y)\right\rangle $ is real, one gets 
\begin{eqnarray}
&&\Delta \left( N_{A}-N_{B}\right) ^{2}  \nonumber \\
&=&\left\langle N_{tot}\right\rangle \left[ 1+4\left( \Delta
p_{o}^{2}-1\right) \left( \int \int_{x>0}u_{o}u_{e}dxdy\right) ^{2}\right]
\label{eq12}
\end{eqnarray}
where $\Delta p_{o}^{2}$ represents the variance of the real part of the odd
mode field fluctuations (normalized so that its value is 1 in vacuum state).
The minimum noise, and therefore the smallest measurable displacement $%
D_{\min}$, is obtained when one uses a perfectly squeezed vacuum state ($%
\Delta p_{o}^{2}=0$). $D_{\min}$ is then equal to zero if $%
\int\int_{x>0}u_{o}u_{e}dxdy=0.5$ \cite{zerobruit}. Given that the two modes
are normalized, one can show that this occurs when $u_{e}$ and $u_{o}$
coincide exactly at all points of one of the two zones, and are opposite in
the other. One notices that in this configuration $u_{e}$ must vary abruptly
from $u_{o}\left( 0,y\right) $to $-u_{o}\left( 0,y\right) $ at the edge
between the two pixels. Such an unusual mode can be experimentally
approximated by inserting a $\pi$ dephasing plate with a sharp edge at $x=0$
in one half of the transverse plane. It must be detected within a very short
distance, as it will diffract quickly when it propagates.\ It can be also
imaged using lenses, with some high spatial frequency filtering because of
the finite size of the optical system. For all these reasons $D_{\min \text{ 
}}$ will not be exactly zero, even with a perfectly squeezed vacuum, but it
will be much smaller than the standard quantum limit.

A simpler configuration, less sensitive to propagation effects, can be
achieved, which consists of a TEM$_{00}$ mode as $u_{e}$, and a perfectly
squeezed vacuum TEM$_{q0}$ mode ($q$ odd, same waist) as $u_{o}$. $D_{min}$
is equal in this case to $0.60D_{sql}$ if $q=1$ and $0.94D_{sql}$ if $q=3$.
This modest improvement with respect to the standard quantum limit is due to
the too slow variation of the odd squeezed mode amplitude when one crosses
the edge $x=0$.\ Much abrupt changes can be obtained by using linear
superpositions of many different transverse TEM$_{pq}$ modes in nonclassical
states.\ In this respect parametric interaction in an optical cavity with a
great number of degenerate transverse modes seems a very promising source of
nonclassical transverse states of light with noise properties varying
abruptly at $x=0$. The quantum properties of the fields generated in
cavities containing thin parametric media, pumped by a plane wave, in the
planar, confocal, or concentric configurations have been recently studied in
great detail, especially below the oscillation threshold.\ In particular the
quasi-planar below threshold configuration has been shown in \cite{gigi3} to
produce a field with almost identical quantum intensity fluctuations at
points symmetrical with respect to the cavity axis.\ This device would
certainly be a good candidate to increase by a large factor the ultimate
sensitivity in the measurement of a small displacement.

In conclusion, we have demonstrated that the standard quantum limit for the
measurement of a small transverse displacement cannot be broken by using
single mode nonclassical states of light, such as those produced in many
experiments performed so far.\ We have shown that different kinds of
multimode non-classical states of light can be used to increase by a large
factor the sensitivity in such a measurement.

We thank L.\ Lugiato, A.\ Gatti and M.\ Kolobov for many enlightening
discussions. This work has been partially funded by an E.C. contract
(ESPRIT\ IV \ ACQUIRE 20029). Laboratoire Kastler Brossel, from Ecole
Normale Sup\'{e}rieure and Universit\'{e} P.M.\ Curie, is associated to CNRS.


\begin{references}
\bibitem{zerobruit}  Actually a perfectly squeezed vacuum is excluded by our
linearization procedure, as it has an infinite noise on the other
quadrature. A simple reasoning shows that $D_{\min }$ is not zero in such a
case, but on the order of $\frac{\Delta }{<N_{tot}>}$.
\end{references}
\end{document}